\documentclass[twocolumn,showpacs,showkeys]{revtex4}
\usepackage{graphicx}
\input{psfig.sty}

\newcommand{\bastar}{\begin{eqnarray*}}
\newcommand{\eastar}{\end{eqnarray*}}
\newskip\humongous \humongous=0pt plus 1000pt minus 1000pt

\newif\ifdtup

\relax
\newcommand{\be}{\begin{equation}}
\newcommand{\ee}{\end{equation}}
\newcommand{\bea}{\begin{eqnarray}}
\newcommand{\eea}{\end{eqnarray}}
\newcommand{\X}{{\vec X}}
\newcommand{\pro}{\partial}
\newcommand{\n}{\hat n}
\newcommand{\oneg}{\displaystyle\frac{1}{g}}

\newcommand{\D}{{\hat D}}

\newcommand{\A}{{\vec A}}
\newcommand{\valpha}{{\vec \alpha}}

\newcommand{\dfrac}{\displaystyle\frac}
\newcommand{\ba}{\begin{array}}
\newcommand{\ea}{\end{array}}

\newcommand{\nn}{\nonumber}
\newcommand{\hn}{\hat n}
\begin{document}
\title  {Electroweak Knot}

\author{Y. M. Cho}
\email{ymcho@yongmin.snu.ac.kr}
\affiliation{School of Physics, College of Natural Sciences, 
Seoul National University,
Seoul 151-742, Korea  \\ 
and \\
C.N. Yang Institute for Theoretical Physics, State University of
New York, Stony Brook, NY 11790, USA}

\begin{abstract}
~~~~~We demonstrate the existence of stable knot solitons in the 
standard electroweak theory whose topological quantum number 
$\pi_3(S^2)$ is fixed by the Chern-Simon index of the $Z$ boson. 
The electroweak knots are made of the helical magnetic 
flux tube of $Z$ boson which has a non-trivial dressing of 
the Higgs field, which could 
also be viewed as two quantized flux rings linked together
whose linking number becomes the knot quantum number. 
We estimate the mass of the lightest 
knot to be around $21~TeV$.
\end{abstract}
\maketitle


Ever since Dirac proposed his theory of monopoles the topological 
objects in physics have been the subject of intensive studies 
\cite{dirac,skyr}. In particular the finite energy topological 
solitons have been widely studied in theoretical physics 
\cite{skyr,thooft}. A remarkable type of solitons is the knots
which have recently appeared almost everywhere,
in nuclear physics in Skyrme theory \cite{fadd1,cho01,cho1},
in plasma physics in coronal loops \cite{fadd2},
in condensed matter physics in one-gap as well as multi-gap
superconductors \cite{cho2,cho3,baba},
and in atomic physics in two-component Bose-Einstein condensates
\cite{cho4,ruo}.
 
Of these, the Faddeev-Niemi knot in Skyrme theory and 
the superconducting knot in ordinary superconductor are 
particularly important.  
The Faddeev-Niemi knot is the prototype of the all the knots, 
which comes from the Skyrme-Faddeev Lagrangian
\bea
&{\cal L}_{SF} = - \dfrac{\mu^2}{2} (\partial_\mu \hat n)^2 -
\dfrac{1}{4} (\partial_\mu \hat n \times \partial_\nu \hat n)^2.
\label{sflag}
\eea
The Lagrangian has the following equation \cite{cho01}
\bea 
&\hn \times
\partial^2 \hn - \dfrac{g}{\mu^2} ( \partial_\mu H_{\mu\nu} )
\partial_\nu \hn = 0, \nn\\
&H_{\mu\nu} = -\dfrac{1}{g}\hn \cdot (\partial_\mu \hn \times
\partial_\nu \hn),
\label{keq}
\eea 
which admits not only the knot
but also the helical baby skyrmion,
a twisted baby skyrmion which is periodic in $z$-coordinate.
The importance of the helical baby skyrmion is that 
the Faddeev-Niemi knot is nothing but 
a vortex ring made of the helical baby skyrmion with 
two periodic ends smoothly connected together. This
tells that the Faddeev-Niemi knot 
originates from the helical baby skyrmion \cite{cho1}. 
The knot is non-Abelian, because the knot topology 
$\pi_3(S^2)$ comes from the $SU(2)$ symmetry of the theory.

On the other hand, the superconducting knot in ordinary
superconductor is an Abelian knot
whose knot topology is fixed by the Chern-Simon index of
the electromagnetic potential \cite{cho2}. Nevertheless this knot is
closely related to the other knots in a fundamental way.
It can also be viewed as a vortex ring made of an helical 
magnetic vortex, the twisted Abrikosov vortex \cite{cho2}.    
This tells that the existence of a helical vortex is
an essential condition for a knot. 
It guarantees the existence of the knot.

{\it The purpose of this Letter is to demonstrate the existence 
of an electroweak knot in Weinberg-Salam theory.
We show that the electroweak knot is made of two quantized
neutral magnetic flux rings linked together, the first one winding
the second $m$ times and the second one winding the first $n$ times,
whose linking number $mn$ becomes the knot quantum number.
Furthermore we predict that the lightest knot has mass
around $21~TeV$ and size of $3.5\times 10^{-18}~m$. 
We also show that the knot has both topological 
and dynamical stability, which comes from the twisted 
topology of the neutral magnetic field.}
If confirmed by experiments, the knot could constitute the first
topological particle in high energy physics.

To establish the existence of the electroweak knot we 
first need to understand the deep connection which exists 
between the Skyrme theory and the non-Abelian 
gauge theory. Let ($\hat n_1,\hat n_2,\hat n$) be 
a right-handed orthonormal basis
in $SU(2)$ space, and consider the following decomposition
of the potential $\vec A_\mu$ into the restricted potential 
$\hat A_\mu$ and the valence potential $\X_\mu$ \cite{cho80,cho81}, 
\bea 
& \vec{A}_\mu =A_\mu \n - 
\oneg \n\times\pro_\mu\n+\X_\mu\nonumber
         = \hat A_\mu + \X_\mu, \nn\\
& A_\mu = \n\cdot \vec A_\mu,
~~~~~\vec{X}_\mu=X_\mu^1~\hat n_1 + X_\mu^2~\hat n_2. 
\label{cdec}
\eea 
Notice that $\hat A_\mu$ is precisely the connection which leaves $\n$ 
invariant under parallel transport, 
\bea 
\D_\mu \n = \pro_\mu \n 
+ g {\hat A}_\mu \times \n = 0. \eea Under the infinitesimal 
gauge transformation \bea \delta \n = - \vec \alpha \times \n  
\,,\,\,\,\, \delta \A_\mu = \oneg  D_\mu \vec \alpha, \eea one has 
\bea &&\delta A_\mu = \oneg \n \cdot \pro_\mu \valpha,\,\,\,\
\delta \hat A_\mu = \oneg \D_\mu \valpha  ,  \nn \\
&&\hspace{1.2cm}\delta \X_\mu = - \valpha \times \X_\mu  . 
\eea 
This tells that $\hat A_\mu$ by itself describes an $SU(2)$ 
connection which enjoys the full gauge degrees of 
freedom. Furthermore $\vec X_\mu$ forms a 
gauge covariant vector field under the gauge transformation. But 
what is really remarkable is that the decomposition is 
gauge-independent. Once $\hat n$ is given, 
the decomposition follows automatically 
independent of the choice of a gauge \cite{cho80,cho81}. 

Notice that $\hat{A}_\mu$ retains the full topological 
characteristics of the original non-Abelian potential. Clearly, 
$\hat{n}$ defines $\pi_2(S^2)$ which 
describes the non-Abelian monopole \cite{cho80m}.  
Besides, with the $S^3$ 
compactification of $R^3$, $\hat{n}$ describes the Hopf 
invariant $\pi_3(S^2)\simeq\pi_3(S^3)$ which characterizes both 
topologically distinct vacua and instanton number \cite{cho79,cho02}. 
Furthermore $\hat{A}_\mu$ has a dual 
structure,
\begin{eqnarray}
& \hat{F}_{\mu\nu} = (F_{\mu\nu}+ H_{\mu\nu})\hat{n}\mbox{,}\nonumber \\
& F_{\mu\nu} = \partial_\mu A_{\nu}-\partial_{\nu}A_\mu \mbox{,}\nonumber \\
& H_{\mu\nu} = -\dfrac{1}{g} \hat{n}\cdot(\partial_\mu
\hat{n}\times\partial_\nu\hat{n}) = \partial_\mu 
C_\nu-\partial_\nu C_\mu,
\end{eqnarray}
where $A_\mu$ and $C_\mu$ are the (chromo)electric and 
(chromo)magnetic potential \cite{cho80,cho81}. 
Notice that $H_{\mu\nu}$ here is exactly the same two-form  
appeared in (\ref{keq}), which admits the potential 
$C_\mu$ because it is closed. 

The decomposition (\ref{cdec}) reveals the deep connection 
between the Yang-Mills theory and 
the Skyrme theory. To see this notice that with
\bea
&\vec H_{\mu\nu}=\partial_\mu \vec
C_\nu-\partial_\nu \vec C_\mu+ g \vec C_\mu \times \vec C_\nu
=H_{\mu\nu}\hat n, \nn\\
&\vec C_\mu= -\dfrac{1}{g}\hat n \times
\partial_\mu\hat n,
\label{ccon}
\eea
we have \cite{cho02}
\bea 
{\cal L}_{SF} = -\dfrac{1}{4} \vec H_{\mu\nu}^2 
- \dfrac{\mu^2} {2} \vec C_\mu^2. 
\eea 
This tells that the Skyrme-Faddeev theory can be interpreted 
as a massive Yang-Mills theory where the gauge potential has the 
special form (\ref{ccon}). This is a first indication that 
the Weinberg-Salam theory could admit a knot
similar to Faddeev-Niemi knot. Our decomposition (\ref{cdec}), which has 
recently been referred to as the ``Cho decomposition'' 
\cite{fadd3,gies,zucc}, plays a crucial role in QCD, in particular
in the calculation of 
the effective action of QCD \cite{cho02c}. 

With these preliminaries we now demonstrate the existence of an 
electroweak knot. Consider the Weinberg-Salam
Lagrangian 
\bea 
&{\cal L} =-\dfrac{1}{4} {\vec F}_{\mu\nu}^2-\dfrac{1}{4}
G_{\mu\nu}^2 -|\tilde D_\mu \phi|^2 
+ m^2\phi^{\dagger}\phi - \dfrac{\lambda}{2} (\phi^{\dagger} \phi)^2, \nn\\
&\tilde D_\mu \phi = ( \partial_\mu + \dfrac{g}{2i} \vec \sigma
\cdot \vec A_\mu +\dfrac{g'}{2i} B_\mu) \phi.
\label{wslag}
\eea 
With the decomposition (\ref{cdec}) we can identify
the Higgs field $\rho$ and $W$ boson $W_\mu$ by, 
\bea 
&\phi =\dfrac{\rho}{\sqrt 2}\xi~~(\xi^{\dagger}\xi =1),
~~~W_\mu = \dfrac{1}{\sqrt 2}(X^1_\mu + i X^2_\mu),
\eea 
and express (\ref{wslag}) in terms of
the physical fields alone
\bea
&{\cal L} = -\dfrac{1}{2}(\partial_\mu \rho)^2
-\dfrac{g^2}{4} \rho^2W_\mu^* W_\mu
-\dfrac{g^2+g'^2}{8} \rho^2 Z_\mu^2 \nn\\
&-\dfrac{\lambda}{8}\big(\rho^2-\rho_0^2\big)^2 
-\dfrac{1}{4} {F_{\mu\nu}^{\rm (em)}}^2 -\dfrac{1}{4} Z_{\mu\nu}^2 \nn\\
&-\dfrac{1}{2}|(D_\mu^{\rm (em)} W_\nu - D_\nu^{\rm (em)} W_\mu)
+ ie \dfrac{g}{g'} (Z_\mu W_\nu - Z_\nu W_\mu)|^2 \nn\\
& +ie F_{\mu\nu}^{\rm (em)} W_\mu^* W_\nu
+ie \dfrac{g}{g'}  Z_{\mu\nu} W_\mu^* W_\nu \nn\\
&+ \dfrac{g^2}{4}(W_\mu^* W_\nu - W_\nu^* W_\mu)^2,
\label{lag1}
\eea
where $\rho_0^2=2m^2/\lambda$, 
$D_\mu^{\rm (em)}=\partial_\mu+ieA_\mu^{\rm (em)}$,
$A_\mu^{\rm (em)}$ and $Z_\mu$ are the electromagnetic 
potential and the $Z$ boson. We emphasize that this expression 
is obtained without any gauge fixing, which is made possible with
the gauge independent decomposition (\ref{cdec}).

Now, with the ansatz
\bea
A_\mu^{\rm (em)} = 0,~~~~~W_\mu = 0,
\eea
we have the following equation
\bea
&\partial^2 \rho - g_0^2 ~Z_\mu^2 ~\rho
= \dfrac{\lambda}{2} (\rho^2 - \rho_0^2) ~\rho, \nn\\
&\partial_\mu Z_{\mu\nu} = j_\nu = g_0^2 ~\rho^2 ~Z_\nu,
\label{zeq}
\eea
where $g_0=\sqrt{g^2+g'^2}/2$ and $Z_{\mu\nu}
=\partial_\mu  Z_\nu-\partial_\nu  Z_\mu$. 
This admits not only a helical vortex but also a knot
made of a twisted magnetic field
of the $Z$ boson.

We construct the helical vortex first.
Choose the cylindrical coodinates $(\varrho,\varphi,z)$
and the following ansatz
\bea
&\rho = \rho(\varrho), \nn\\
&Z_\mu= \dfrac{1}{g_0} \big(n Z_1(\varrho) \partial_\mu\varphi
+ mk Z_2(\varrho) \partial_\mu z \big).
\label{hvans}
\eea
With this we have
\bea
&Z_{\varrho\varphi} = \dfrac{n}{g_0} \dot Z_1,
~~~~Z_{\varrho z} = \dfrac{mk}{g_0} \dot Z_2,
~~~~Z_{\varphi z} = 0, \nn\\
&j_\mu = g_0 \rho^2 \Big(n Z_1 \partial_\mu\varphi
+ mk Z_2 \partial_\mu z \Big),
\label{hvew}
\eea
which clearly shows that both the magnetic flux and the supercurrent
have a helical structure.

With the ansatz (\ref{zeq}) is reduced to
\bea
&\ddot{\rho}+\dfrac{1}{\varrho}\dot\rho
- \Big(\dfrac{n^2}{\varrho^2} Z_1^2
+ m^2 k^2 Z_2^2 \Big)\rho 
= \dfrac{\lambda}{2}(\rho^2-\rho_0^2)\rho, \nn\\
&\ddot{Z}_1-\dfrac{1}{\varrho}\dot{Z}_1 -g_0^2 ~\rho^2
Z_1 = 0, \nn\\
&\ddot{Z}_2+\dfrac{1}{\varrho}\dot{Z}_2 -g_0^2 ~\rho^2 Z_2 = 0.
\label{hveq}
\eea
Now, we impose the following boundary condition
\bea
&\rho (0) = 0,~\rho(\infty) = \rho_0,
~~Z_1 (0) = 1,~Z_1 (\infty) = 0.
\label{hvbc}
\eea
As for $Z_2$ we require that the vortex carries 
a non-vanishing supercurrent. This uniquely fixes the boundary
condition for $Z_2$, which dictates that $Z_2(\infty)=0$ 
but $Z_2(0)$ must have a logarithmic divergence.

With this we obtain the helical vortex shown in Fig.1.
Notice that, when $m=0$, the solution (with $Z_2=0$) describes
the well-known $Z$ boson vortex \cite{vach}. But when $m$ is not zero,
it describes a helical vortex which has a non-vanishing supercurrent
(not only around the $z$-axis but also) along the $z$-axis,
\bea
&i_{\hat z} = mkg_0 \dfrac{}{} \int \rho^2 Z_2 
\varrho d\varrho d\varphi
= -\dfrac{2\pi mk}{g_0} (\varrho \dot Z_2) \Big|_{\varrho=0}.
\label{sc}
\eea
This confirms that the logarithmic divergence of $Z_2$
at the origin is what we need to make the vortex 
a superconducting string. But here, of course, 
the supercurrent is neutral, not electromagnetic.

\begin{figure}
\includegraphics[scale=0.7]{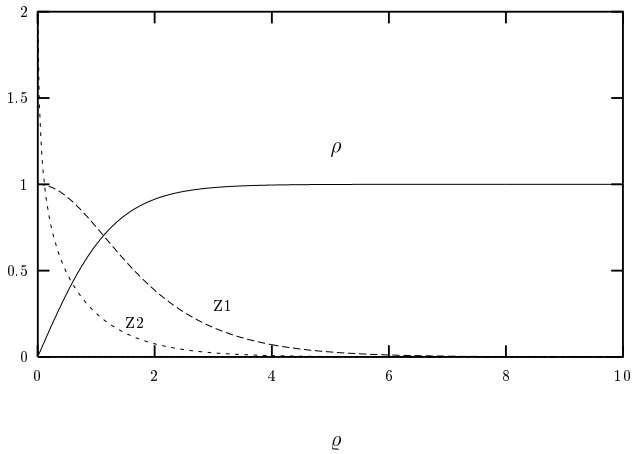}
\caption{The helical electroweak vortex with 
$m=n=1$ in Weinberg-Salam theory. Here we have put $g_0=1,~\lambda=2$,
$k=\rho_0/10$, and $\varrho$ is in the unit of $1/\rho_0$.}
\label{ewknot}
\end{figure}

Clearly the vortex has the magnetic field $H_{\hat z}$
and the quantized magnetic flux $\Phi_{\hat z}$ along the $z$-axis
\bea
H_{\hat z}= \dfrac{n}{g_0} \dfrac{\dot Z_1}{\varrho},
~~~\Phi_{\hat z}= \dfrac{}{} \int H_{\hat z} \varrho d\varrho d\varphi
= \dfrac{2\pi n}{g_0}.
\label{zflux}
\eea
But it also has the magnetic field $H_{\hat \varphi}$ 
around the vortex
\bea
H_{\hat \varphi} = \dfrac{mk}{g_0} \dot Z_2.
\label{rflux}
\eea
Unfortunately this produces an infinite
the magnetic flux $\Phi_{\hat \varphi}$ around
the vortex, because of the singularity at the origin.
So one might conclude that the helical vortex is unphysical
which does not exist, because one need an infinite energy to
create it.

The importance of the helical vortex, however, is not 
in that it is physical but in that it ensures the existence
of a knot. To see this notice that with the vortex 
we can make a vortex ring by smoothly bending and connecting 
two periodic ends. In this vortex ring the infinite magnetic flux 
$Z_2$ becomes finite because the finite supercurrent (\ref{sc}),
in the vortex ring, should produce a finite magnetic flux
passing through the disk of the ring. 
This tells that the infinite magnetic flux
$\Phi_{\hat \varphi}$ of the vortex is an artifact of 
a straight vortex, which disappears in the vortex ring. 
Furthermore, we can certainly make the finite magnetic flux
passing through the knot disk to have the quantized value $2\pi m/g_0$
by adjusting the supercurrent of the ring with $k$.
Remarkably, this vortex ring now becomes a topologically
stable knot. To see this notice that 
the magnetic flux of the knot is made of two flux rings,
a $2\pi m/g_0$ flux around the knot tube and a $2\pi n/g_0$ 
flux along the knot. Moreover the two flux rings
are linked together, whose linking number becomes $mn$.
This is precisely the mathematical description of a knot,
two rings linked together. This assures that the vortex ring
does indeed become a topological knot. 
The knot quantum number is described by
the Chern-Simon index of $Z_\mu$ field,
\bea
&Q = \dfrac{g_0^2}{32\pi^2} \int \epsilon_{ijk} Z_i Z_{jk} d^3x
=mn,
\label{ewkqn}
\eea
which describes the non-trivial topology $\pi_3(S^2)$ of
the magnetic field $Z_{\mu\nu}$. Notice that this 
is formally identical to the
quantum number of the Faddeev-Niemi knot \cite{cho01,cho1}.
The only difference is that here the (chromo)magnetic potential
$C_\mu$ is replaced by $Z_\mu$. Furthermore, just as
in the Faddeev-Niemi case, the Chern-Simon index is given by
the linking number of two magnetic fluxes. 

Obviously the two flux rings linked together can not be
separated with a continuous deformation of the field configuration.
This provides the topological stability
of the knot.

Furthermore, this topological stability of the knot
is backed up by a dynamical stability. 
This is because the supercurrent along the knot
now generates a net angular momentum 
around the knot which naturally provides the centrifugal 
repulsive force to prevent the collapse of the knot. 
This tells that the knot is dynamically stable.
Another way to understand the dynamical stability
is to notice that, when the knot tries to shrink,
the energy density of magnetic field trapped in
the knot disk increases inevitably.
This generates a repulsive force against the collapse,
which makes the knot stable. 

We emphasize hat the stability
of the knot crucially depends on the helical structure 
of the supercurrent and magnetic field. 
Without this there is neither the topological 
stability nor the dymamical stability.

We can estimate the mass of the electroweak knot. 
To do this notice that in the absence of the Higgs
field the knot becomes almost identical to the Faddeev-Niemi knot
in Skyrme theory. From this observation we may estimate
the energy density (per length) and the energy, thus the radius  of 
the knot as \cite{cho4,ussr} 
\bea 
&{\cal E} = \pi n \rho_0^2,
~~~E \geq 16\pi^2 \times 3^{3/8} ~(mn)^{3/4}~g_0\rho_0, \nn\\ 
&R = \dfrac{E}{2\pi {\cal E}} 
\simeq \dfrac{8g_0 \times 3^{3/8}}{\rho_0}
\dfrac{m^{3/4}}{n^{1/4}}.
\label{ken}
\eea 
So, with the experimental values $g_0 \simeq 0.36$ 
and $\rho_0 \simeq 246~GeV$ we expect
the lightest electroweak knot to have mass around 
$21~TeV$ and radius of about $3.5\times 10^{-18}~m$, with tube size
$0.7\times 10^{-18}~m$.  

Clearly the electroweak knot is closely related to 
the Faddeev-Niemi knot. But one can not overemphasize the striking
similarity between the electroweak knot and 
superconducting knot in ordinary superconductor \cite{cho2}.
Mathematically they are identical. The only difference 
is that the superconducting knot exists at the atomic scale
and carries the real electric supercurrent, whereas
the knot here exists at the electroweak scale and carries 
the neutral current. It is really remarkable that mathematically
identical knots can exist in totally different physical
enviornments, in $eV$ scale and in $TeV$ scale.

We believe that our analysis has established
the existence of a electroweak knot
beyond reasonable doubt. Of course, one might like to see
an analytic solution of the knot. Unfortunately this is 
impossible. At present even the simplest knot
does not allow an analytic solution \cite{fadd1}.
But one can construct an approximate solution analytically
which has all the charactristic features of 
the electroweak knot, using the toroidal coordinates and adopting 
an educated ansatz \cite{cho2,cho5}. This provides another
evidence which endorses the fact that the electroweak knot 
is for real. A challenging task now would be 
to confirm the existence of the electroweak knot by 
high energy experiments. We hope that the LHC at CERN could
confirm the existence of the electroweak knot. 

The physical implications of the electroweak knot
and the details of our analysis will be published separately \cite{cho5}.

{\bf ACKNOWLEDGEMENT} 

~~~We thank Professor C. N. Yang for the illuminating 
discussions. The work is supported by the Basic Research Program of
Korea Science and Engineering Foundation (Grant R02-2003-000-10043-0)
and by the BK21 project 
of the Ministry of Education.

\end{document}